# Density of Nanometrically Thin Amorphous Films Varies by Thickness

*Yael Etinger-Geller,* [†, ‡] *Alex Katsman,* [†] *and Boaz Pokroy* [†, ‡, *]

†Department of Materials Science and Engineering, Technion − Israel Institute of Technology, 3200003 Haifa, Israel, ‡Russell Berrie Nanotechnology Institute, Technion Israel Institute of Technology, 3200003 Haifa, Israel

ABSTRACT: Organisms in nature can alter the short-range order of an amorphous precursor phase, thereby controlling the resulting crystalline structure. This phenomenon inspired an investigation of the effect of modifying the short-range order within the amorphous phase of a selected material. Amorphous thin films of aluminum oxide deposited by atomic layer deposition method were found to vary structurally as a function of size. Thinner films, as predicted and also confirmed by atomistic simulations, exhibited more 4-coordinated Al sites. These atomistic alterations were expected to change the amorphous thin film's average density. The density indeed varied with the alumina layer thickness, and the measured effect was even stronger than predicted theoretically. This effect is explained in terms of the deposition process, where each newly deposited layer is a new surface layer that 'remembers' its structure, resulting in thin films of substantially lower density.

**INTRODUCTION**

Amorphous materials are important for a number of different applications in science and technology[1] owing to their unique electronic, optical, and mechanical properties.[2] Although such materials are in common use, scientists have only recently begun to explore their extraordinary structures.[3] Whereas crystalline materials are characterized by a periodic and predictable atomic arrangement, in amorphous materials the order decays rapidly with the distance.[3, 4] It is nevertheless possible to describe the structure of amorphous materials in terms of short-range order (coordination number, nearest neighbors, bond length, bond angles), which is similar around atoms of the same kind, and has a typical bond distance of up to 2−3 atoms.[3, 5] Fine changes in the atomistic structure can lead to new, fascinating phenomena, most of which are not yet known.

The inspiration for this study comes from nature, where amorphous phases characteristically have various advantages and play significant roles.[6] One such characteristic is the ability to serve as a transient precursor phase for controlled mineralization into a specific crystalline structure, even if this is not the thermodynamically preferred one.[7, 8] This is achieved by controlling of the short-range order in the amorphous precursor so that it resembles that of the desired crystalline polymorph.[7] Such control is achieved in nature via different additives, such as polymeric molecules or magnesium ions, which become incorporated into the crystal and induce precipitation of a specific phase.[6-8]

Finding a way to emulate this manipulative technique synthetically would have a profound impact, as many technological applications utilize different characteristics of amorphous phases.[9]

Moreover, in many cases a specific crystalline polymorph is required for a particular function[10] and indeed, many studies were conducted over the years so as to achieve controlled crystallization. [11-14]

We chose to utilize size effects in order to alter the atomistic structure in the amorphous state.[15] Size effects have been extensively studied in crystalline materials and found to change various properties, both structural[16, 17] and functional,[18, 19] in nanosized materials. These effects, which arise from highly pronounced surface stress and energy, are not restricted to crystalline materials but can exist in any solid substance. Furthermore, amorphous materials, like crystalline ones, can exist in different solid structures,[20] implying the possibility of switching between them. We chose atomic layer deposition (ALD) as our sample preparation technique, as this is an excellent method of producing high-quality, conformal, and pinhole-free thin films of different systems.[21-25]

In the quest to control the short-range order within an amorphous phase, our group recently performed a breakthrough study in which size effects were found to alter the short-range order in amorphous $Al_2O_3$ nanofilms.[15] The results indicated that the surface of an amorphous $Al_2O_3$ nanofilm is characterized by a short-range order that is richer in 4-coordinated Al sites ($Al_4$ sites)[26] than that of the bulk structure. Thus, the thinner the amorphous film, the more its short-range order resembles that near the surface. These results are also supported by atomistic simulations.[27] We therefore expect that the dependence of the short-range order on size in amorphous $Al_2O_3$ will yield, as in crystalline materials, different size-dependent physical properties.

In this study we investigated the size dependence of density in amorphous $Al_2O_3$ nanofilms. We discuss the experimental results in the framework of the developed energetic model, taking into account surface reconstruction driven by a decrease in surface energy.

**EXPERIMENTAL SECTION**

Thin films of $Al_2O_3$ were deposited on p-type Si (100) wafers by the use of thermal ALD (ALD R-200 Advanced tools, Picosun, Finland). Our selected working temperatures were 200°C and 350°C, as this process has a wide ALD window[28] allowing conformal and precise deposition. Silicon wafers were rinsed with ethanol and dried under a $N_2$ gas stream prior to the deposition processes. Trimethylaluminum (TMA) was used as the precursor and $H_2O$ as the oxidizer. The system was operated under a continuous flow of $N_2$ carrier gas (99.999% pure). A basic ALD cycle consisted of 0.1 s TMA pulse (room temperature), 6 s $N_2$ purge, 0.1 s $H_2O$ pulse (room temperature) and 6 s $N_2$ purge. One ALD cycle yields one monolayer of substance,[28] and by repetition of the basic cycle the thickness of the film can be strictly controlled. The same conditions were used for the growth of all samples. The thickness of the films was measured by direct ellipsometry of the wafer.

To examine the variations in density we chose the x-ray reflectometry (XRR) technique, performed using the x-ray diffractometer (SmartLab, Rigaku, Japan). N is a surface-sensitive characterization method that is used to study thin films, and from which different parameters such as the thickness, roughness and density of the thin film can be measured[29]. Electronic density is proportional to the square root of the critical angle,[10] from which the physical density[30] of a thin layer can be calculated by simulations.

**RESULTS**

To study the size effect on the density of the amorphous $Al_2O_3$ structure we used ALD to deposit nanofilms of $Al_2O_3$ of varied thicknesses, as the size effect is expected to be more pronounced in the thinner films. Using this technique we achieved good linear growth with excellent control of the thickness, and a growth rate per cycle of approximately 1Å at both working temperatures, 200°C and 350°C. The deposited films were of high quality, conformal, and pinhole free (Figure 1)

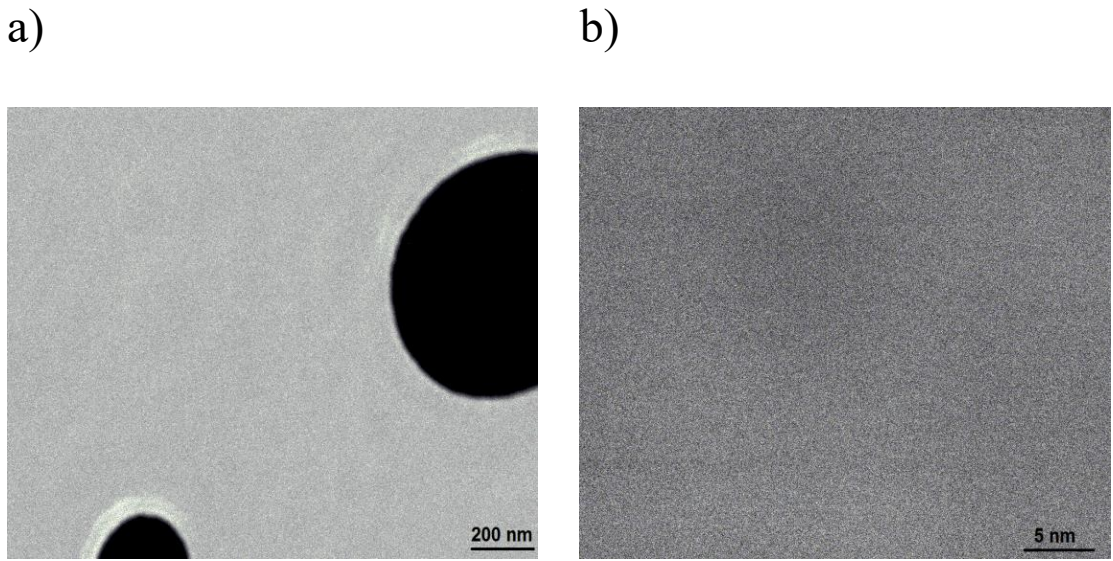

**Figure 1.** High-angle annular dark field scanning transmission electron microscopy (HAADF STEM) measurements of a 20-nm thick amorphous $Al_2O_3$ layer deposited on a holey carbon TEM grid.

Samples of three different thicknesses were scanned by X-ray reflectivity (XRR) (Figure 2A). An XRR spectrum is characterized by its periodic pattern, from which the film's thickness, density and roughness can be extracted. The thickness of the film is proportional to the frequency of the intensity fluctuations[29], and the spectra obtained from thicker samples are indeed seen to be characterized by higher fluctuation frequency. In addition, the extended spectra are indicative of the high quality and low roughness of the deposited films.

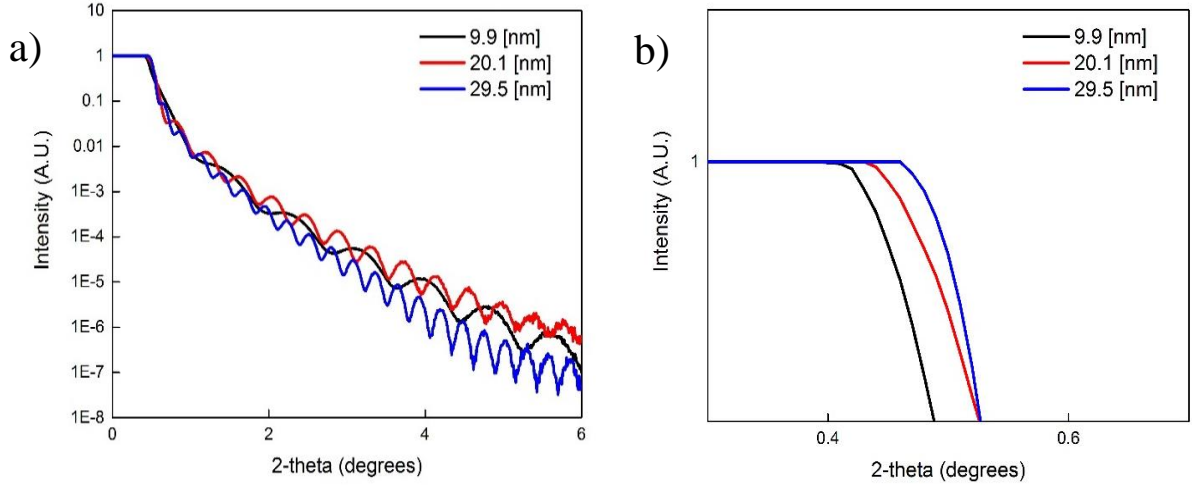

**Figure 2.** (a) XRR measurements of amorphous $Al_2O_3$ samples of different thickness. (b) Shift in the critical angle with size.

Another parameter that can be extracted from the XRR spectra is the density. The density, $\rho$, is determined by the critical angle $\theta_c$, which is proportional to the square root of the density according to equation (1)[31]:

$$\theta_c = \sqrt{\left(\frac{r_e \lambda^2}{\pi}\right) N_0 \rho \sum_i x_i (Z_i + f_i') \Big/ \sum_i x_i M_i} \qquad (1)$$

where $r_e$ is the classical radius of an electron, $N_0$ is the Avogadro number, $\lambda$ is the x-ray wavelength, and $Z_i$, $M_i$, $x_i$ and $f'_i$ are the atomic number, atomic weight, atomic ratio and atomic scattering factor of the i-th atom, respectively.

The critical angle range (Figure 2B) shows a shift of the critical angle towards higher angles for the thicker samples, indicative of their higher density. To examine the variation in critical angle, we performed slow high-resolution scans on samples of different thicknesses. The critical angle was defined as the angle at which, owing to refraction of the x-ray beams, the reflected intensity starts to decrease. [29]

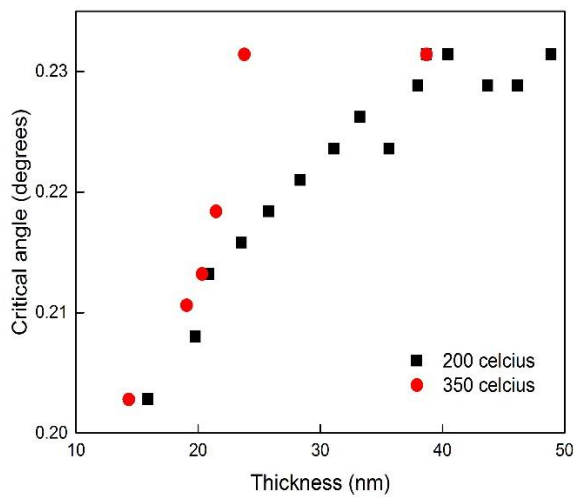

**Figure 3.** Dependence of the critical angle on thickness, in thin films deposited at 200°C (black) and 350°C (red).

The results presented in Figure 3 indicate that the critical angle depends significantly on the thickness of the amorphous thin film. The ALD working temperature has only little influence on the critical angle (and on the density) of very thin films (below ~20 nm), but 'saturation' of the critical angle with increasing thickness is achieved faster at a higher working temperature.

**DISCUSSION AND THEORETICAL MODELING**

The size dependence of the density of very thin films can be assumed to be associated with an increase in the total free energy owing to the presence of two interfaces. Let's estimate the possible change in the film's density that occurs because of a change in the average interatomic interaction energy. The stable interatomic distance in alumina corresponds to the minimum value of its total interatomic potential energy, including that of the two-body O−Al, O−O and Al−Al parts and the three-body Al−O−Al and O−Al−O parts.[32] In amorphous alumina, the major part of the interatomic interactions is represented by the O−Al bonds. Let's assume, for simplicity, that the O−Al interatomic potential, $V_{O-Al}$, determines the average interatomic distance in amorphous alumina. This is given by the expression[32]:

$$V_{O-Al} = \frac{H_{O-Al}}{r^9} + \frac{Z_O Z_{Al}}{r} e^{-r/\lambda} - \frac{D_{O-Al}}{r^4} e^{-r/\xi} - \frac{W_{O-Al}}{r^6} \qquad (2)$$

where $H_{O-Al}$ = 249.3eV·Å$^9$, $D_{O-Al}$ = 50.15eV·Å$^4$, $W_{O-Al}$ = 0, $Z_{Al}$ = 1.5237e, $Z_O$ = −1.02e, e = 1.6·10$^{-19}$C, $\lambda$ = 5Å, and x = 3.75Å.

The minimum value of the potential energy (2) corresponds to an O−Al distance of 1.6Å (see supplementary Note I), which is indeed close to the distance between an oxygen atom and its nearest tetrahedral void ($r^{O-tetrahedral\ void}$ = 1.68Å).

The average O−Al interatomic potential energy in a thin Al$_2$O$_3$ film can be written as following (see supplementary Note II):

$$V_{O-Al} \cong V^{\infty}_{O-Al} + \frac{\Delta E}{N}, \qquad (3)$$

where N is the number of atomic oxygen monolayers, $V^{\infty}_{O-Al}$ is the O−Al interatomic potential in the bulk, and $\Delta E/N$ is the change in this potential due to the presence of two external surfaces. The energy of unrelaxed external surfaces can be quite a large ($\gamma^{unrel}$~10 J/m$^2$)[33], and it decreases substantially during relaxation ($\gamma^{rel}$~1 J/m$^2$). The major part of this energy is presumably distributed in the bulk of the film during reconstruction of the near-surface atomic layers, and this is achieved through the decrease in the average interatomic interaction energy, in accordance with equation (3). The reduced interatomic bonds correspond to the increased interatomic distances.

From equation (2) we can extract the average O−Al distance as an inverse function of the interatomic potential $V_{O-Al}$:

$$\bar{r}_m = f^{-1}(V_{O-Al}(N)) \equiv \psi(N) \tag{4}$$

By these means, using equation (3), we can determine its dependence on the thickness. The change in average density can be written as following:

$$\frac{\rho - \rho_0}{\rho_0} = \left(\frac{\bar{r}_{m0}}{\bar{r}_m}\right)^3 - 1 = \frac{\psi^3(0)}{\psi^3(N)} - 1 \tag{5}$$

where $\bar{r}_{m0}$ is the average O−Al distance corresponding to a very thick film. The calculated dependence of the relative density change as a function of the film thickness, with the use of interatomic potential (2), is presented in Figure 4.

For $\bar{r}_m > \bar{r}_{m0}$, the potential energy is well approximated by the Coulomb term (since the other terms diminish rapidly).

In this case, using equation 3, one can obtain:

$$\frac{\bar{r}_{m0}}{\bar{r}_m} \exp\left(\frac{\bar{r}_{m0} - \bar{r}_m}{\lambda}\right) \approx 1 + \frac{\Delta E}{V_{O-Al}^\infty N}. \tag{6}$$

The change in density for the case $(\bar{r} - \bar{r}_{m0}) \ll \lambda$ can be approximated by means of the following equation:

$$\frac{\rho - \rho_0}{\rho_0} \approx \left(\frac{\bar{r}_{m0}}{\bar{r}_m}\right)^3 - 1 \approx \frac{3\Delta E}{V_{O-Al}^\infty N} = \frac{1.56 \gamma_s d_{111}^2}{V_{O-Al}^\infty N}. \tag{7}$$

Assuming $V^\infty_{O-Al} = -8.3\,eV$, $\gamma_s = 9.45\,J/m^2$, and $d_{111} = 0.224$ nm, yields:

$$\frac{\rho - \rho_0}{\rho_0} \approx -\frac{1.7}{N}. \tag{8}$$

Therefore, the relative change in average density is expected to be inversely proportional to the number of atomic monolayers, which is, of course, directly related to the thickness of the film (Figure 4). Experimentally, the change in density of the amorphous alumina thin films can be found via XRR measurements, using equation (1):

$$\frac{(\rho_0 - \rho)}{\rho_0} = 1 - \left(\frac{\theta_c}{\theta_{c,0}}\right)^2 \tag{9}$$

where $\theta_{c,0}$ is the critical angle for the thickest films, which also have the largest density, $\rho_0$. The experimental results are compared with the theoretically predicted dependence, equation (8), in Figure 4.

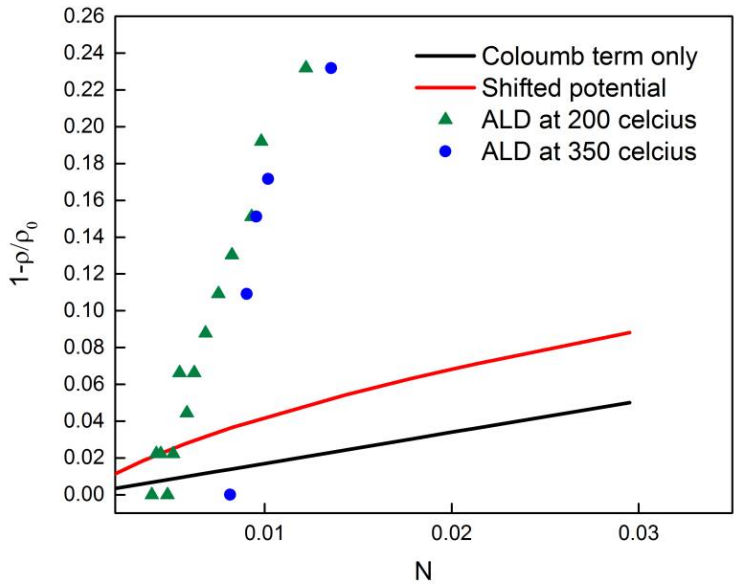

**Figure 4.** Comparison between the theoretical investigation and the experimental results.

A near-linear relationship between the relative density change and the inverse film thickness is indeed observed (triangle and circle dots in Figure 4). The experimental effect, however, is much stronger than expected from the above model. The relative density change as large as (0.2 ÷ 0.25) obtained for films (15 ÷ 20) nm thick cannot be explained only by energetic factors such as surface energy relaxation; the latter predicts such high density changes only for very thin films with N ≤ 7 (see equation (8)). Additional thermal treatment of the low-density films at various temperatures ranging from 300°C to 750°C had no measurable impact on the critical angle. The ALD working temperature did not influence the appearance of low-density very thin films, while the density 'saturation' rate increased with increasing working temperature. These findings suggest that metastable low-density amorphous alumina layers are formed in the near-surface regions of relatively large thickness (15−20 nm). Those layers are probably a result of the combined effect of the surface layer relaxation and the kinetics of the ALD process, as discussed below.

The most superficial layer can be viewed as an outcome of the dynamic reconstruction of several outer atomic layers during continuous deposition of alumina by the ALD method. To illustrate such possible reconstruction, let's consider three near-surface parallel (111) atomic layers in an ideal γ-alumina structure, two of which—A and C—contains only oxygen atoms (terminating the surface A-layer and the inner C-layer), and the intermediate layer B consists of only $Al_4$ and $Al_6$ atoms, as shown schematically in Figure 5. Every oxygen atom in the close-packed layer (A or C) has 3 nearest-neighbor oxygen atoms in the second oxygen layer. Every Al atom in the octahedral position has 3 neighbor oxygen atoms in every oxygen layer, while an Al atom in the tetrahedral position has 3 neighbor oxygen atoms in one layer and 1 in another. Every Al atom thus has, on average, 2 oxygen neighbors in each layer. The distance between the layers corresponds to a zero total interaction force acting on every oxygen layer in the direction perpendicular to it. In approximation of nearest neighbor pair interaction, this force can be expressed as following:

$$F_1 = n_s^{Ox} \left[ \frac{4}{3} x \left. \frac{\partial V_{Al-O}}{\partial r} \right|_{r_{tet}} (\vec{n}_{111}\vec{n}_{tet}) + 2(1-x) \left. \frac{\partial V_{Al-O}}{\partial r} \right|_{r_{oct}} (\vec{n}_{111}\vec{n}_{oct}) + 3 \left. \frac{\partial V_{O-O}}{\partial r} \right|_{r_{O-O}} (\vec{n}_{111}\vec{n}_{O-O}) \right] = 0 \quad (10)$$

where $x$ is the fraction of Al4 atoms ($x = 0.3$ for ideal γ-alumina); $\vec{n}_{tet}$ and $\vec{n}_{oct}$ are unit vectors from tetrahedral and octahedral voids, respectively, to the nearest oxygen atom; $\vec{n}_{O-O}$ is the unit vector in the direction between two nearest oxygen atoms in the neighbor layers; and $\vec{n}_{111}$ is the unit vector in the [111] direction. For the fcc γ-alumina $(\vec{n}_{111}\vec{n}_{tet}) = 2/3$, $(\vec{n}_{111}\vec{n}_{oct}) = 1/\sqrt{3}$, $(\vec{n}_{111}\vec{n}_{O-O}) = \sqrt{2/3}$, $r_{tet} = a\sqrt{3}/4$, $r_{oct} = a/2$, and $r_{O-O} = a/\sqrt{2}$ and $d_{111} = a/\sqrt{3}$. Solving equation (10) by substituting the interatomic potentials[32] values into the equation, yields the lattice parameter a = 0.371 nm, which seems a rather good approximation of the real γ-alumina structure ($a_\gamma$ = 0.388 nm).

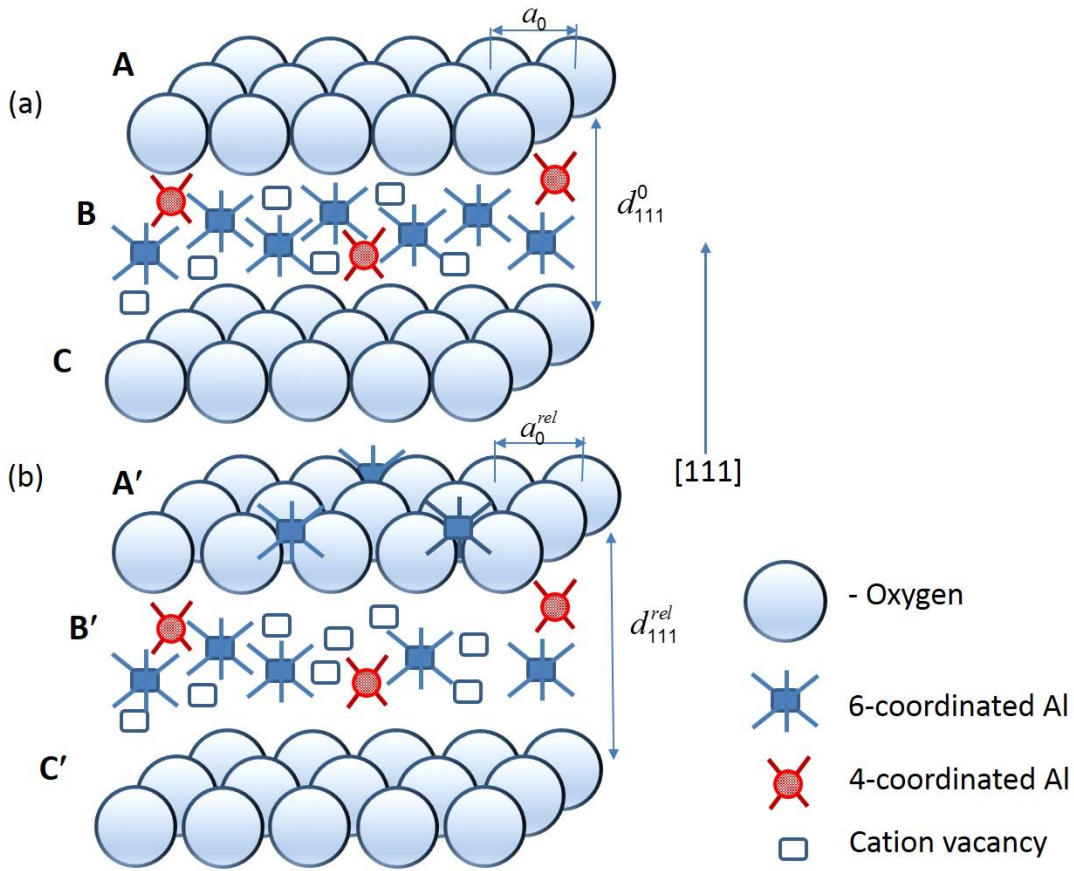

**Figure 5.** Schematic structure of the near-surface atomic layers in alumina. (a) Unrelaxed ideal structure of γ−Al$_2$O$_3$; (b) Reconstructed structure.

During reconstruction of the surface, $Al_6$ atoms may pass to the outer layer A in order to bond with the non-bridging oxygen atoms, and will probably take the triangular void positions in the close-packed oxygen layer to form a new relaxed outer layer A'. This process is dictated by a reduction in total interaction energy of the near-surface structure.[34] The vacancies left by the $Al_6$ atoms can be occupied, at least partially, by other Al atoms from the inner Al layer (not shown here) during structure relaxation.

The interaction energy increases (thus becoming more negative) with the decreasing fraction of $Al_6$ atoms in the structure. As a result, the fraction of $Al_4$ atoms increases. At the same time, the equilibrium distance between the layers will increase due to the Coulomb terms in the potential[32]: the attractive Al−O forces between the layers will decrease, while new repulsive Al−Al forces between the B' and A' layers will appear. The change in interlayer distance depends on the fraction of Al atoms ($z$) passing to the outer A' layer and the fraction of $Al_6$ atoms ($y$) remaining in the B' layer after relaxation. The interlayer equilibrium distance is a function of $y$ and $z$ and determined by the following equation:

$$x \frac{\partial V_{Al-O}}{\partial r}\bigg|_{r_{tet}} + y \frac{\sqrt{27}}{4} \frac{\partial V_{Al-O}}{\partial r}\bigg|_{r_{oct}} + \frac{3}{4} zxp \frac{\partial V_{Al-Al}}{\partial r}\bigg|_{d_{111}/2} + \sqrt{\frac{243}{32}} \frac{\partial V_{O-O}}{\partial r}\bigg|_{r_{O-O}} = 0 \quad (11)$$

where p = 0.1 is the probability of finding an Al atom in a tetrahedral position in the B layer. The variation in total interlayer binding energy with lattice parameter, and the numerical solution of equation (11) for different values of y and z, are given in supplementary Note III. The lattice parameter increases and the effective density decreases with increasing $Al_6$ atoms fraction passing to the outer layer. The relative changes in density may reach magnitudes of about 0.3 and larger for z = 0.3 ÷ 0.4 (see Figure 8 in supplementary Note III).

The transfer of $Al_6$ atoms and their embedding in the outer oxygen layer should trigger subsequent reconstruction of the A as well as of the B and C layers. Such reconstruction might result in destruction of the crystal structure of the layers, a process that can be viewed as amorphization of the near-surface region. Computer simulations[34] with modified Born-Mayer-Huggins potentials have also shown that the reconstruction of (111) surfaces of γ-alumina is extensive and that the near-surface regions become amorphous, or at least significantly distorted, at a depth of about 0.7 nm.

Therefore, a change in the density of amorphous alumina films can be accompanied by a change in the average coordination number of Al atoms. A molecular dynamics study of amorphous alumina[34] showed that the change in density from 3.0 to 3.3 g/cm$^3$ results in a substantial change of short-range order. Moreover, experimental investigation of the size effect on the short-range order in a nanosized amorphous alumina[15] confirmed that the fraction of tetrahedral Al sites is greater in thinner amorphous films. The pronounced effects on the presence of Al$_4$ atoms were observed in films less than 20 nm thick. But the intriguing question remains: how can the reconstruction of a very narrow near-surface layer (only ≈ 0.7−1.0 nm thick) result in a low-density layer of 15−20 nm?

The main feature of the ALD process is the following: each newly deposited atomic layer in turn becomes an outer surface layer. Accordingly, several formerly outer atomic layers will now constitute a near-surface layer with an amorphous structure characterized by a higher fraction of Al$_4$ sites and low density. This layer, although by now an inner layer, 'remembers' its surface structure for a comparatively long time. The formation of relatively thick and stable low-density layers (15−20 nm) may lead to the occurrence of a new metastable amorphous phase of low density with an enlarged fraction of Al$_4$ sites. Reconstruction and densification of this layer into a regular amorphous phase might be restrained for various reasons, in particular owing to a high activation-energy barrier to further reconstruction. An increase in density where the thickness is more than ~15 nm signifies such phase transformation in the depth of the film. The dependence of the density 'saturation' rate on the ALD working temperature exemplifies a kinetic factor that determines the friable/dense amorphous/amorphous phase transformation.

## CONCLUSIONS

The near-surface-layer reconstruction driven by the decrease in surface energy, combined with the ALD kinetic process in which every newly deposited atomic layer appears—albeit transiently—as the most superficial layer, results in the formation of a metastable, low-density, amorphous alumina structure. Thus, the density of ALD produced thin amorphous alumina films remains significantly lower than that of bulk amorphous alumina with thicknesses of up to 40−50 nm and enlarged fractions of $Al_4$ atoms. The average short-range order parameter, being a function of the film density, conforms to the thickness of the film. The lower density thus corresponds to a smaller coordination number of aluminum atoms.

## ASSOCIATED CONTENT

Supporting information available: Note I, stable interatomic distance of Al-O potential interaction. Note II, average O-Al interatomic potential energy in a thin $Al_2O_3$ layer. Note III, variation of interlayer interaction energy and interlayer distance as a function of $Al_6$-atom fractions in A' and B layers. This material is available free of charge *via* the internet at http://pubs.acs.org.

**AUTHOR INFORMATION**

**Corresponding Author**

*E-mail: bpokroy@tx.technion.ac.il

**Author Contributions**

All authors contributed to writing of the manuscript, and all approved its final version.

.‡These authors contributed equally. (match statement to author names with a symbol)

**Funding Sources**

**Notes**

Any additional relevant notes should be placed here.

**ACKNOWLEDGMENT**

The research leading to these results has received funding from the European Research Council under the European Union's Seventh Framework Program (FP/2013-2018) / ERC Grant Agreement n. 336077. We are grateful to Dr. Oleg Kreinin for his help with preparing the samples and operating the ALD system. We are also grateful to Dr. Yaron Kauffmann for his help with TEM experiments.

**ABBREVIATIONS**

ALD, atomic layer deposition; STEM, scanning transmission electron microscopy; XRR, x-ray reflectometry.